\documentclass[twocolumn]{aastex631}

\usepackage{todonotes}
\usepackage{mathtools}
\usepackage{graphicx}	
\usepackage{amsmath}	

\newcommand{\dd}{{\rm d}}
\newcommand{\myr}{\mbox{${\rm Myr}$}}
\newcommand{\gyr}{\mbox{${\rm Gyr}$}}
\newcommand{\pc}{\mbox{${\rm pc}$}}
\newcommand{\msun}{\mbox{M$_\odot$}}

\defcitealias{Webb2019a}{Paper I}

\begin{document}

\title{A systematic analysis of star cluster disruption by tidal shocks -- II.~Predicting star cluster dissolution rates from a time-series analysis of their tidal histories}

\shorttitle{Cluster disruption for arbitrary tidal histories}
\shortauthors{Webb, J.J. et al.}

\author{Jeremy J. Webb}
\affiliation{Department of Science, Technology and Society, Division of Natural Science, York University, 4700 Keele St, Toronto ON M3J 1P3, Canada}
\affiliation{Department of Astronomy and Astrophysics, University of Toronto, 50 St. George Street, Toronto, ON, M5S 3H4, Canada}

\author{Marta Reina-Campos}
\affiliation{Canadian Institute for Theoretical Astrophysics (CITA), University of Toronto, 60 St George St, Toronto, M5S 3H8, Canada}
\affiliation{Department of Physics \& Astronomy, McMaster University, 1280 Main Street West, Hamilton, L8S 4M1, Canada}

\author{J.~M.~Diederik Kruijssen}
\affiliation{Technical University of Munich, School of Engineering and Design, Department of Aerospace and Geodesy, Chair of Remote Sensing Technology, Arcisstr.~21, 80333 Munich, Germany}
\affiliation{Cosmic Origins Of Life (COOL) Research DAO, coolresearch.io}






\begin{abstract}

Most of the dynamical mass loss from star clusters is thought to be caused by the time-variability of the tidal field (``tidal shocks''). Systematic studies of tidal shocks have been hampered by the fact that each tidal history is unique, implying both a reproducibility and a generalisation problem. Here we address these issues by investigating how star cluster evolution depends on the statistical properties of its tidal history. We run a large suite of direct $N$-body simulations of clusters with tidal histories generated from power spectra of a given slope and with different normalisations, which determine the time-scales and amplitudes of the shocks, respectively. At fixed normalisation (i.e.\ the same median tidal field strength), the dissolution time-scale is nearly independent of the power spectrum slope. However, the dispersion in dissolution time-scales, obtained by repeating simulations for different realisations of statistically identical tidal histories, increases with the power spectrum slope. This result means that clusters experiencing high-frequency shocks have more similar mass loss histories than clusters experiencing low-frequency shocks. The density-mass relationship of the simulated clusters follows a power-law with slope between $1.08$ and $1.45$, except for the lowest normalisations (for which clusters effectively evolve in a static tidal field). Our findings suggest that star cluster evolution can be described statistically from a time-series analysis of its tidal history, which is an important simplification for describing the evolution of the star cluster population during galaxy formation and evolution.

\end{abstract}



\section{Introduction} \label{sec:intro}

As star clusters evolve within a galactic potential, they are driven towards dissolution due to a wide range of internal and external processes \citep{heggie03, Gieles2011}. Integrated over a star cluster's lifetime, the dominant external mass loss mechanism is posed by the time-variability of the tidal field, i.e.\ so-called ``tidal shocks'' \citep{lamers06, elmegreen10, elmegreen10b, kruijssen15b, miholics17, pfeffer18, li18,ReinaCampos2022}. These tidal shocks accelerate the stars in a cluster, causing them to reach energies in excess of the binding energy and thus allowing them to escape, which can alter both the mass and structure of a star cluster \citep{spitzer58, ostriker72, chernoff86, aguilar88, chernoff90, kundic95, gieles06, kruijssen11, gieles16, MartinezMedina2022}. In  smooth tidal fields, shocks occur for clusters with non-circular orbits due to perigalactic passes \citep{gnedin97, gnedin99, baumgardt03, webb13, webb14a} and disc passages \citep{gnedin97, kruijssen09, webb14b}. In more realistic tidal fields that contain substructure, shocks can occur due to interactions with giant molecular clouds \citep{gieles06, lamers06}, dark matter substructure \citep{Webb2019b}, spiral arms \citep{gieles07}, and galaxy merger-induced structures \citep{kruijssen12, Renaud2013, Mamikonyan17}. 

Including shocks in models of cluster evolution has been shown to be essential in reproducing the observed distribution of cluster ages in the Solar Neighbourhood. This Galactic region was originally found to be devoid of open clusters older than $\sim1$~Gyr by \citet{Oort1958} and \citet{Wielen1971}, to the extent that stellar evolution, two-body relaxation and tidal stripping alone are not sufficient to explain the dearth of old clusters \citep{Lamers2005}. This tension was relieved with the inclusion of the effects of tidal shocks due to giant molecular clouds in models of cluster evolution  \citep[e.g.][]{lamers06,gieles06}. Only when including shocks do old clusters get destroyed sufficiently quickly in the Solar Neighbourhood and the relatively young cluster ages are reproduced.

Given the prevalence and impact of tidal shocks throughout a cluster's lifetime, it is extremely important to develop an understanding of how shocks affect cluster evolution. Such an understanding is essential for using observations of present-day clusters as cosmological tools to study the formation, evolution and structure of their host galaxy \citep{Forbes2018,kruijssen19b}. In \citet{Webb2019a} (\citetalias{Webb2019a} of this series), we provide a thorough discussion of how previous studies have developed a framework for modelling the effect of tidal shocks on star clusters \citep{spitzer58, aguilar85, aguilar86, gnedin03, gieles06, prieto08, kruijssen11, gieles16}. In terms of recent work in this area appearing since \citetalias{Webb2019a}, we highlight the paper by \citet{MartinezMedina2022}, who simulate the evolution of star clusters that are perturbed by either point-masses or disc passages, and identify quantitative differences between shock-driven mass loss as a result of these differences in geometry. 

The above body of work on modelling the effect of tidal shocks on star clusters has shown that the relative amount of mass loss generated by a tidal shock is proportional to the relative energy gain, i.e.:
\begin{equation}
    \frac{\Delta M}{M_0} = f\frac{\Delta E}{E_0} ,
\end{equation}
where $\Delta M$ is the change in cluster mass due to the shock, $M_0$ is the cluster mass before the shock, $f$ is a proportionality constant that depends on the shock geometry, $\Delta E$ is the change in cluster energy due to the shock, and $E_0$ is the binding energy of the cluster before the shock. The change in cluster energy depends on the properties of the shock and the cluster as
\begin{equation}
    \frac{\Delta E}{E_0} = \frac{\Delta E_{\rm imp}}{E_0} A_{\rm w}(x),
\end{equation}
where $\Delta E_{\rm imp}$ is the energy change expected for an impulsive tidal shock, $x\equiv \tau/t_{\rm dyn,h}$ is the ratio between the shock duration $\tau$ and the dynamical time at the half-mass radius $t_{\rm dyn,h}$, and $A_{\rm w}(x)$ is an ``adiabatic correction'' introduced by \citet{spitzer87} to account for shocks that are slow compared to the dynamical times of the cluster stars, which makes the shock adiabatic in nature rather than impulsive \citep{weinberg94a, weinberg94b, weinberg94c, gnedin97, gnedin99, gnedin03}. The adiabatic correction is given by \citep{gnedin99}:
\begin{equation}\label{eqn:adiabatic}
A_{\rm w}(x) = \left(1+x^2\right)^{-3/2}.
\end{equation}

In order to account for the density profile of the cluster before the shock, \citet{MartinezMedina2022} add a new parameter $\epsilon$ to the adiabatic correction, such that
\begin{equation}
A_{\rm w}(x) = \left(1+\epsilon^2x^2\right)^{-3/2}.
\end{equation}

For a given shock strength, the authors fit for different values of $\epsilon$ that depend on the cluster's central density parameter $W_0$, and they successfully reproduce the change in energy of the cluster. However, the authors note that this formalism does not apply to point-mass perturbations. Tidal shocks due to disc passages are completely compressive, whereas point-mass shocks have equal compressive and extensive components. Thus, in the case of shocks due to point masses, the change in the energy of the cluster is instead found to increase as a function of ${\tau}/{t_{\rm dyn,h}}$  \citep{MartinezMedina2022}. 

Despite several decades of work dedicated to studying the effect of tidal shocks on star cluster evolution, a dynamically-motivated framework that accurately reproduces how clusters are affected by shocks in $N$-body simulations remains elusive. For a given cluster mass and density profile, it would be ideally possible to predict the evolution of the cluster directly from its tidal history. This prediction would be given by the time-evolution of the tidal tensor,
\begin{equation}
    T_{ij}(t) = -\dfrac{\dd^2 \Phi(t)}{\dd x_i \dd x_j},
\end{equation}
which describes the time-variation of the tidal field generated by the external potential, $\Phi(t)$ that is experienced by the cluster. The tidal history defines a tidal heating parameter \citep{prieto08}:
\begin{equation}
   I_{\rm tid} = \sum_{i,j} \left(\int T_{ij} dt\right)^{2}A_{{\rm w},ij}(x),
\end{equation}
which serves as an indicator of how much energy is injected into a cluster during a given encounter. This parameter can be used to determine the disruption time-scale of the cluster due to shocks \citep{prieto08, kruijssen11}. The time evolution of the tidal tensor, $T_{ij}$, can be extracted from large-scale cosmological simulations in order to simulate the evolution of star clusters affected by a realistic, time-dependent potential \citep[e.g.][]{renaud11, pfeffer18, Rodriguez2022, ReinaCampos2022}.

Systematic studies of how tidal shocks impact star cluster evolution in direct $N$-body simulations are necessary to make further progress in this field. However, efforts in this direction have been hampered by the fact that each tidal history is unique, because they are generated by the specific structures that exist in the direct environment of the cluster. Hence there are a vastly larger number of degrees of freedom compared to (now classical) studies of star cluster disruption in smooth tidal fields \citep[e.g.][]{baumgardt03}, and implies both a reproducibility and a generalisation problem:
\begin{enumerate}
    \item Unless data are shared, no two research groups can simulate cluster evolution for the same tidal history. When considering results for different tidal histories, it is not clear how these should be compared.
    \item The results of a single simulation for a single tidal history are meaningless when it is not possible to determine how representative the simulation is for star clusters with other properties or subjected to other tidal histories.
\end{enumerate}
These fundamental problems may be addressed if it is possible to reduce the dimensionality of the problem, for instance, by identifying quantities that can describe tidal histories in a statistical sense and at the same time can be used to predict the disruption time-scales and mass loss rates of clusters evolving in these tidal histories.

In \citetalias{Webb2019a}, we take a first step towards a systematic analysis of star cluster disruption due to tidal shocks, by subjecting model star clusters to idealized tidal shocks. More specifically, we modelled extensive shocks in $T_{xx}$ as step functions of a given duration, while the rest of components were kept at zero. The main findings of \citetalias{Webb2019a} are as follows:
\begin{enumerate}
    \item The amount of mass lost during a shock primarily depends on the strength of the shock and the cluster's density within the half-mass radius.
    \item A pair of shocks has the same effect on a cluster as a single shock with the same $I_{\rm tid}$ as long as the time between shocks is less than the cluster's crossing time.
    \item A pair of shocks has a different effect on a cluster than a single shock with the same $I_{\rm tid}$ if the time between shocks is greater than the cluster's crossing time due to cluster evolution between shocks.
    \item Correcting classic tidal shock theory \citep[e.g.][]{spitzer58,spitzer87,prieto08} to account for the escape time-scale of unbound stars results in a dynamically-motivated prediction for how clusters lose mass due to tidal shocks that matches the $N$-body simulations.
\end{enumerate}

The next step towards a complete understanding of how tidal shocks affect star clusters involves moving beyond the step function description of shocks and exploring how strongly the dissolution of clusters depends on the details of individual shocks, or if this process can be characterised by treating tidal histories as statistical ensembles of shocks. In this study, we build on the work done in \citetalias{Webb2019a} by performing 540 different $N$-body simulations of star cluster disruption. The simulations subject clusters to complex tidal histories that are generated from power spectra with given slopes and normalisation factors. The slopes of the power spectra govern the frequency at which the cluster is subjected to shocks and the normalisation factor sets the strength of both the the mean background tidal field \footnote{The normalisation sets the mean background tidal field such that only extensive shocks are being considered as the tidal tensor is always positive.)} and the shocks themselves. By generating multiple tidal histories with the same statistical properties (i.e.\ with the same power spectrum slope and normalisation factor), it becomes possible to compare the evolution of star clusters with tidal histories that are unique but are statistically indistinguishable. This comparison will allow us determine whether cluster evolution depends strongly on the details of individual shocks or predominantly on the total energy injected. In addition, it will help address the reproducibility and generalisation problems of star cluster evolution with time-variable tidal histories that we outline above.

Establishing a relationship between a cluster's evolution and the statistical properties of its tidal history, as opposed to the specific form of its tidal history, is a step towards being able to model cluster evolution without the need for a detailed representation of the cluster's external environment. If instead it can be established that the cluster's orbit and the distribution of matter within its galactic environment will result in a tidal history characterized by a given power spectrum and normalisation factor, then a realistic time-dependent tidal heating parameter can be generated and incorporated into analytic models for cluster evolution \citep[e.g.][]{pfeffer18,Hui2019}. This would unlock the ability to sample star cluster evolution from a distribution of mass loss and density evolutionary tracks, as opposed to requiring the direct modelling of the millions of star clusters that have existed throughout the history of a Milky Way-mass galaxy. By reducing complex tidal histories to two metrics describing the shape of their power spectrum, it also enables systematic comparisons between simulation suites developed by different groups. This approach will strengthen our ability to study how clusters evolve in different galactic environments.

The final step will be to express the metrics that describe the tidal history of a cluster in terms of the properties of the cold interstellar medium (ISM) through which it traverses. Such a translation step is beyond the scope of this paper and is deferred to future work, but we mention it here specifically to illustrate how this experiment contributes to a future in which the evolution of a star cluster population can be tied directly to the properties of the galactic environment. For example, sub-mm observations of the cold ISM in nearby galaxies would provide sufficient information to set the power spectrum slope and normalization factor, allowing one to model the evolution of the star cluster population in such an environment, without needing to know individual cluster orbits or the detailed ISM structure. Comparing the demographics of the resulting cluster population would unlock end-to-end experiments of our understanding of star cluster formation and evolution.

This paper is organised as follows. In Section \ref{sec:nbody}, we outline the details of our direct $N$-body simulations and explain how individual tidal histories are generated. The dissolution times and structural evolution of model clusters are compared in Section \ref{sec:results}. We summarize our findings in Section \ref{sec:conclusion}.

\section {$N$-body Simulations} \label{sec:nbody}

In order to compare the evolution of star clusters with different tidal histories, we make use of the direct $N$-body code \texttt{NBODY6tt} \citep{aarseth03, renaud11}. With this code, we simulate star clusters with a range of masses and sizes (i.e.\ and thus, densities). \texttt{NBODY6tt} is a modified version of \texttt{NBODY6} \citep{aarseth03} that allows for a time-dependent tidal tensor to represent the external tidal field experienced by the cluster \citep{renaud11}.

Similar to \citetalias{Webb2019a}, our main suite of simulations consist of star clusters that initially contain $50,000$ stars of mass $0.6~\msun$ and have a half-mass radius of $8~\pc$. Initially, the density profile of each cluster is a Plummer sphere \citep{plummer11}. We also consider clusters containing $1/3$ and $1/10$ of the fiducial number of stars, to assess whether cluster mass affects the impact of the same tidal history. For each of the low-mass clusters, we consider cases where the clusters have either the same density or the same size as our fiducial models. In almost all cases, simulations are run with force calculation time-steps equal to $0.003~\myr$. In cases where the cluster's dissolution time is greater than $1~\gyr$, the time-step sizes are doubled to $0.006~\myr$ for computational efficiency. We have ensured that this change in the time-step  has no effect on the cluster's dissolution time.

Next, we address the reproducibility and generalisation problems of star cluster evolution with time-variable tidal histories that we defined in Section~\ref{sec:intro}, by generating these histories from a simple power spectrum that is defined by only two parameters. Specifically, we generate histories for the $T_{xx}$ component of the tidal tensor using the fast Fourier transform library of functions within \texttt{numpy} \citep{Walt2011}. We define the power spectrum, $P(k)\propto k^{-\alpha}$, between frequencies $k=0.005$--$1.0~\myr^{-1}$, and we consider three different slopes of $\alpha =\{1, 2, 3\}$. A phase between $0$ and $2\pi$ is randomly sampled from a uniform distribution and then the absolute value of the one-dimensional inverse discrete Fourier transform of each power spectrum is determined. We take the absolute value to ensure that we are only considering extensive shocks. In this study, the one-dimensional inverse discrete Fourier transform is initially normalized by $1/10$ and then set equal to $T_{xx}$. Values of $T_{xx}$ are provided at $0.01~\myr$ time-steps. This initial normalisation ensures that the mean tidal field strength is comparable to star cluster formation environments \citep[e.g.][]{pfeffer18,li22,meng22}, which results in most of our clusters dissolving within $1~\gyr$. Histories generated from power spectra with slopes $\alpha=1$ and $\alpha=3$ are further multiplied by factors of $\sim 4.9$ and $\sim 0.1$, respectively to ensure that the median tidal field strength is the same as for histories generated with slopes $\alpha=2$. For a given slope, we also consider a range of tidal field strengths by re-normalising the tidal histories by factors of $S=\{0.3, 0.6, 1, 1.7, ,3\}$. Beyond $T_{xx}$, all other components of the tidal tensor are set to zero. 

\begin{figure}
    \includegraphics[width=\hsize]{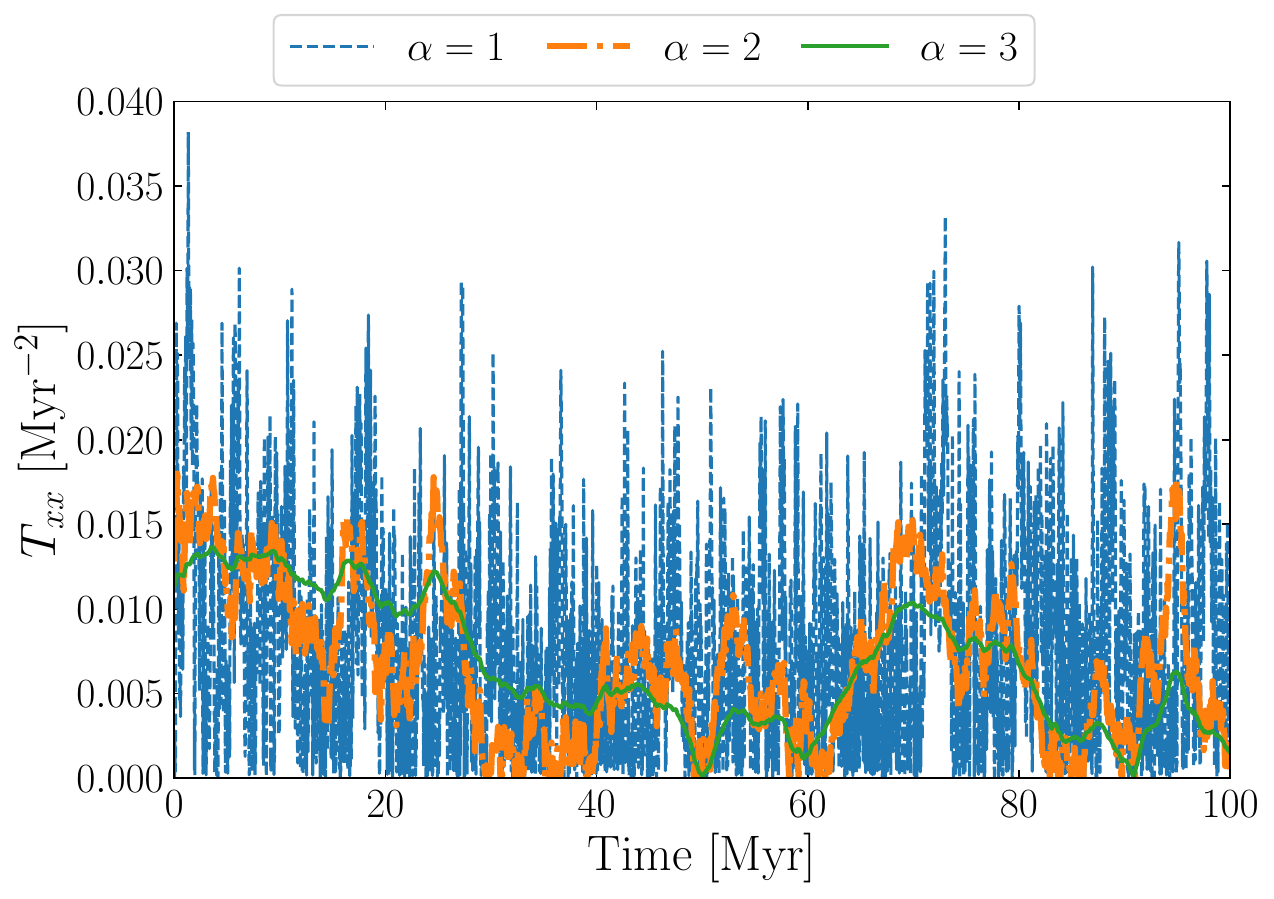}
    \caption{The $T_{xx}$ component of the tidal tensor as a function of time, generated from power spectra of slopes $\alpha=\{1,2,3\}$, as indicated in the legend.}
    \label{fig:ttplot}
\end{figure}

Figure \ref{fig:ttplot} illustrates three tidal histories generated from power spectra with slopes of $\alpha=\{1, 2, 3\}$ and adopting the fiducial normalisation of $S=1$. For the tidal history generated from a slope of $\alpha=1$, clusters experience a large number of short-duration shocks. Decreasing the slope to $\alpha=3$ places more power at low frequencies and thereby decreases the number of shocks experienced by a cluster while increasing the durations of the shocks. By changing the normalisation $S$, we are simply increasing or decreasing the strength of the shocks. 

\begin{figure}
    \includegraphics[width=\hsize]{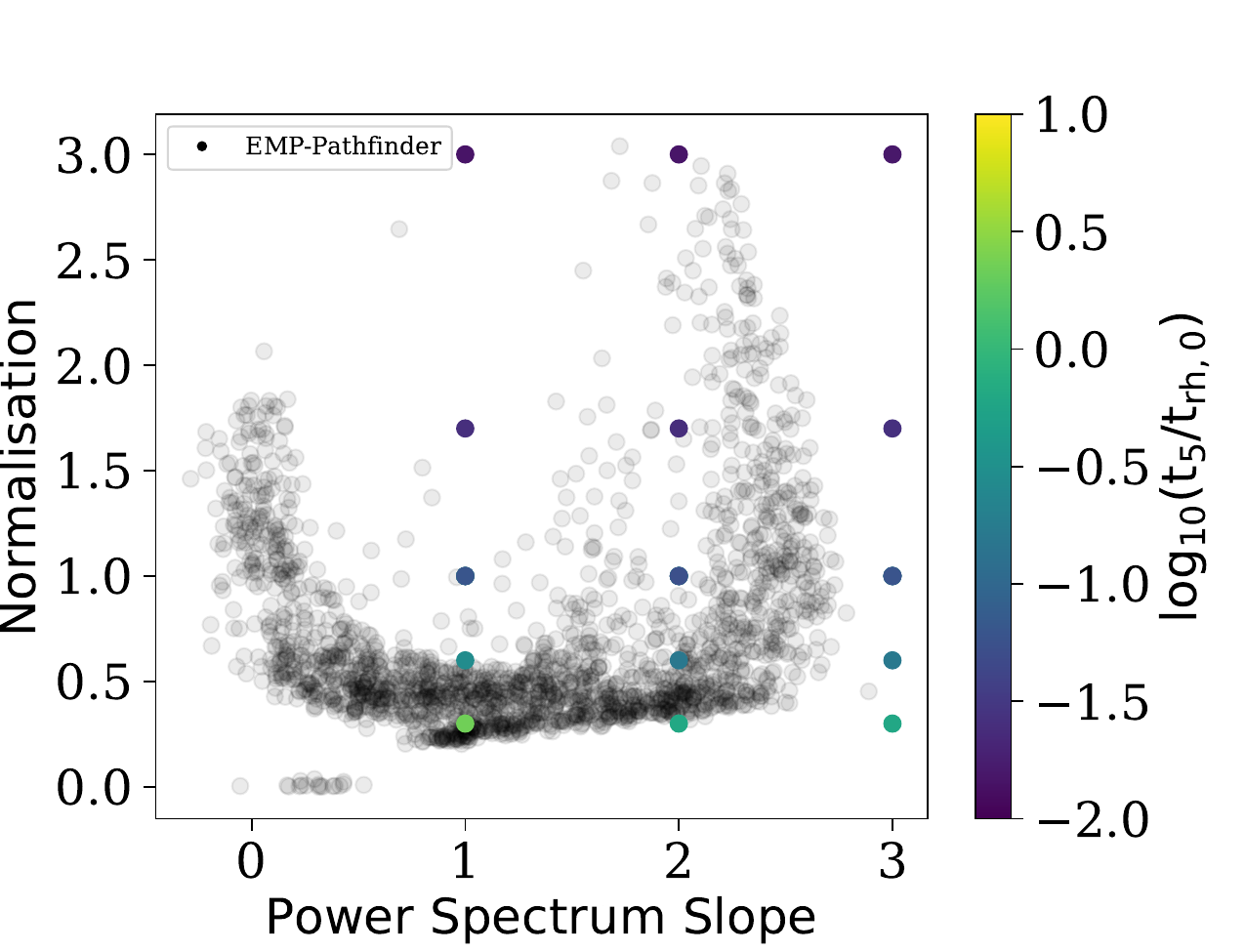}
    \caption{Parameter space used to define the power spectra from which the tidal histories are generated in this work. The data points indicate the combinations of power spectrum slope and tidal history normalisation, with colours indicating the mean cluster dissolution time-scale in units of the initial half-mass relaxation time for the fiducial clusters. The colour scale highlights that all simulations except the one with $\{\alpha,S\}=\{1,1/3\}$ have dissolution time-scales shorter than their initial half-mass relaxation time, indicating that they are disrupted by tidal shocks. For comparison purposes, the black points show the corresponding normalisation and power spectrum slope of random star cluster tidal histories extracted from EMP-\textit{Pathfinder} \citep{ReinaCampos2022}. The tidal histories used in our suite of simulations spans the same parameter space covered by cosmologically-motivated tidal histories.}
    \label{fig:grid}
\end{figure}

In Table~\ref{table:models}, we provide the complete suite of simulations used in this study, with Figure \ref{fig:grid} illustrating the power spectrum slopes, tidal history normalisations, and the ratio of the mean 5~per~cent dissolution time (defined to correspond to the time at which only 5~per~cent of the initial mass is left) to initial relaxation time for the fiducial models. The names of the models in Table~\ref{table:models} are based on the initial number of stars in the cluster (`N'), the initial cluster half-mass radius (`R'), the slope of the power spectrum (`A'), and the tidal history normalisation (expressed through a scale factor `S'). For each combination of power spectrum slope, tidal history normalisation, cluster mass, and cluster size, 20 simulations have been run with different realisations of statistically identical tidal histories, for a total of 540 simulations. Exploring differences between a given a set of 20 simulations allows for an exploration of how variations in tidal histories with the same statistical properties affect cluster evolution.

We illustrate the normalisation and power spectrum slopes covered by the tidal histories used in our suite of simulations in  Figure \ref{fig:grid}. This figure also compares our suite to tidal histories extracted from the suite of cosmologically zoom-in simulations of Milky Way-mass galaxies from EMP-\textit{Pathfinder} \citep{ReinaCampos2022}. These simulations model the concurrent formation and evolution of star cluster populations alongside their host galaxies over a Hubble time in a sub-grid fashion. A key aspect is that they incorporate the physics required to model the multi-phase nature of the ISM, which is shown to be the critical ingredient needed to accurately reproduce how star clusters evolve for more than $10~\gyr$. We randomly select the tidal history experienced by 2000 star particles in the simulation MW04. With them, we first determine the time evolution of the largest eigenvalue of the tidal tensor, and we then calculate the power spectrum of their absolute value. The power spectrum slope is determined by fitting a linear relation in logarithmic space to the resulting power spectrum. By randomly sampling the star particles, we select tidal histories that are experienced from $\sim 12~\gyr$ to a mere few tens of Myr. The corresponding normalisation values of the EMP-\textit{Pathfinder} histories are determined by comparing the median value of each tidal history's maximum eigenvalue to the median value of our tidal histories with known normalisations.

Directly comparing our synthetic tidal histories to cosmologically-motivated tidal histories, we see that the two sets of tidal histories span the same normalisation-power spectrum slope parameter space for power spectrum slope values greater than or equal to $\alpha \geq 1.0$. EMP-\textit{Pathfinder} has a sub-population of histories where the measured power spectrum slope is less than $\alpha \leq 1.0$, however these histories correspond to cases where the tidal field is static and are not applicable to a study of cluster disruption via tidal shocks.

While the strength and time-variation in our tidal histories are comparable to the maximum eigenvalues of tidal histories extracted from EMP-\textit{Pathfinder}, it remains important to remember that our generated histories are not physical in that they do not represent interactions between a cluster and substructure. Our tidal tensors are all one dimensional, with only $T_{xx}$ being assigned positive values based on our method for generating tidal histories from a power spectrum. Hence the shocks are extensive and only occur along the $x$-axis, with the positive trace of the tidal tensor implying a negative mass density. Similar to \citetalias{Webb2019a}, using such idealised tensors allow for a systematic study of how star clusters are affected by tidal shocks since the number of free parameters in the experimental setup have been minimized. By first establishing a relationship between cluster evolution and the global properties of these simplified tidal histories, in this case governed by two parameters, we form a basis for expanding this approach to more complex tidal histories.

\begin{table*}
\centering
\begin{tabular}{lcccc}
Name  & Initial number of Stars & Initial half-mass radius [pc] & Power Spectrum Slope & Tidal history normalisation \\\hline\hline
N50\_R8\_A1\_S03 & 50,000 & 8 & 1 & 0.3 \\
N50\_R8\_A1\_S06 & 50,000 & 8 & 1 & 0.6 \\
N50\_R8\_A1\_S10 & 50,000 & 8 & 1 & 1 \\
N50\_R8\_A1\_S17 & 50,000 & 8 & 1 & 1.7 \\
N50\_R8\_A1\_S30 & 50,000 & 8 & 1 & 3 \\
N50\_R8\_A2\_S03 & 50,000 & 8 & 2 & 0.3 \\
N50\_R8\_A2\_S06 & 50,000 & 8 & 2 & 0.6 \\
N50\_R8\_A2\_S10 & 50,000 & 8 & 2 & 1 \\
N50\_R8\_A2\_S17 & 50,000 & 8 & 2 & 1.7 \\
N50\_R8\_A2\_S30 & 50,000 & 8 & 2 & 3 \\
N50\_R8\_A3\_S03 & 50,000 & 8 & 3 & 0.3 \\
N50\_R8\_A3\_S06 & 50,000 & 8 & 3 & 0.6 \\
N50\_R8\_A3\_S10 & 50,000 & 8 & 3 & 1 \\
N50\_R8\_A3\_S17 & 50,000 & 8 & 3 & 1.7 \\
N50\_R8\_A3\_S30 & 50,000 & 8 & 3 & 3 \\ \hline
N5\_R8\_A1\_S10 & 5,000 & 8 & 1 & 1 \\
N5\_R8\_A2\_S10 & 5,000 & 8 & 2 & 1 \\
N5\_R8\_A3\_S10 & 5,000 & 8 & 3 & 1 \\
N5\_R3\_A1\_S10 & 5,000 & 3.5 & 1 & 1 \\
N5\_R3\_A2\_S10 & 5,000 & 3.5 & 2 & 1 \\
N5\_R3\_A3\_S10 & 5,000 & 3.5 & 3 & 1 \\
N16\_R8\_A1\_S10 & 16,666 & 8 & 1 & 1 \\
N16\_R8\_A2\_S10 & 16,666 & 8 & 2 & 1 \\
N16\_R8\_A3\_S10 & 16,666 & 8 & 3 & 1 \\
N16\_R5\_A1\_S10 & 16,666 & 5.2 & 1 & 1 \\
N16\_R5\_A2\_S10 & 16,666 & 5.2 & 2 & 1 \\
N16\_R5\_A3\_S10 & 16,666 & 5.2 & 3 & 1 \\
\end{tabular}
\caption{\label{table:models} Summary of the models considered. The columns list the name of the model, the initial number of stars, the initial cluster half-mass radius, the slope of the power spectrum used to generate the cluster's tidal history, and the normalisation of the tidal history. The models in the first half of the table have the same number of stars and initial sizes and are considered to be the fiducial models. The models in the second half of the table have fewer stars than the fiducial models, but have either the same density within the half-mass radius or the same half-mass radius. For each combination of parameters, 20 different tidal histories were generated and 20 different simulations run to dissoluition. Most clusters dissolve within 1 Gyr, but the N50\_R8\_A1\_S03 models take nearly 5 Gyr to dissolve.}
\end{table*}

\section{Results} \label{sec:results}

\subsection{Cluster mass evolution}

We have simulated a range of star clusters with tidal histories that have been randomly generated from power-law power spectra and have been normalised to different mean tidal field strengths. The purpose of these simulations is to determine whether cluster evolution strongly depends on the exact details of its tidal history (i.e.\ the strength and duration of individual tidal shocks) or if it can be approximated based on the statistical properties of its tidal history (i.e.\ the power spectrum shape, which in turn determines the mean background tidal field strength, mean shock strength, and typical shock duration). To explore the effects of the uniqueness of a cluster's tidal history on its evolution, we generate 20 random tidal histories from each power spectrum, defined by the power spectrum slope and tidal history normalisation. 

\begin{figure}
    \centering
    \includegraphics[width=\hsize]{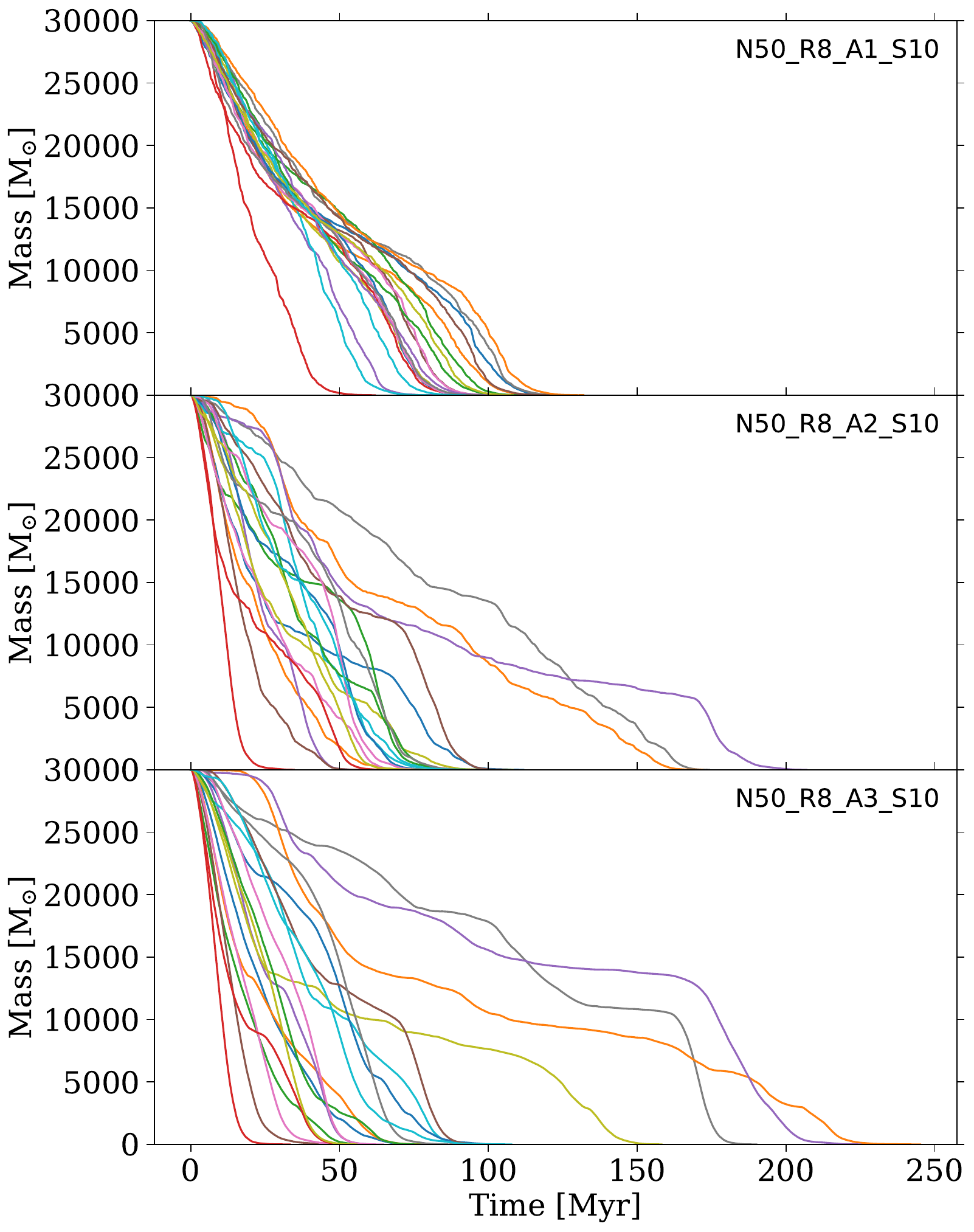}
    \caption{Cluster mass as a function of time for tidal histories drawn from power spectra with slopes of $\alpha=1$ (\textit{top panel}), $\alpha=2$ (\textit{middle panel}), and $\alpha=3$ (\textit{bottom panel}). Each panel shows the mass evolution for 20 different realisations of the tidal history. For comparison purposes, we include in each panel the evolution of a model cluster subject to a static tidal field of strength equal to the median tidal field strength of the 20 realisations (dashed black line). In all panels, this `no-shocks' model leads to a lifetime that is considerably longer than that of most other models in that panel.}
    \label{fig:mplot}
\end{figure}

Figure \ref{fig:mplot} illustrates the mass evolution of 20 star clusters with tidal histories generated from power spectra with different slopes. Within a given panel, the dissolution time-scales vary between clusters with different realisations of their tidal histories. If a cluster happens to experience a strong shock early in its lifetime, it may dissolve faster than a cluster that does not receive a strong shock until late in its lifetime. Additionally, the range of evolutionary channels is narrower for N50\_R8\_A1\_S10 models (top panel) than for the N50\_R8\_A2\_S10 and N50\_R8\_A3\_S10 models (bottom panel), i.e.\ tidal histories generated from a flatter power spectrum (implying higher-frequency variations of the tidal field) result in more similar cluster evolution across different realisations. This means that the degree to which a cluster's evolution depends on the uniqueness of its tidal history depends on the slope of the power spectrum from which the tidal history is generated.

For comparison purposes, we also compare the evolution of the models to a model cluster that experiences a static tidal field in Fig.~\ref{fig:mplot}. The strength of the static tidal field is set equal to the median strength of the non-static tidal field experienced by the N50\_R8\_A1\_S10, N50\_R8\_A2\_S10, and N50\_R8\_A3\_S10 models. In the simulations considered here, clusters that experience a static tidal field take 1.7 times longer to dissolve than the median of the identical clusters subjected to statistically similar tidal shock histories.

To better understand and quantify how the statistical properties of the tidal history affect cluster evolution, we consider each model cluster's dissolution time and structural evolution in Sections \ref{sec:dissolution} and \ref{sec:density}. We further explore how these results depend on the initial mass and size of the cluster itself in Section \ref{sec:mass}.

\subsection{Star cluster dissolution times} \label{sec:dissolution}

\begin{figure*}
    \centering
    \includegraphics[width=\hsize]{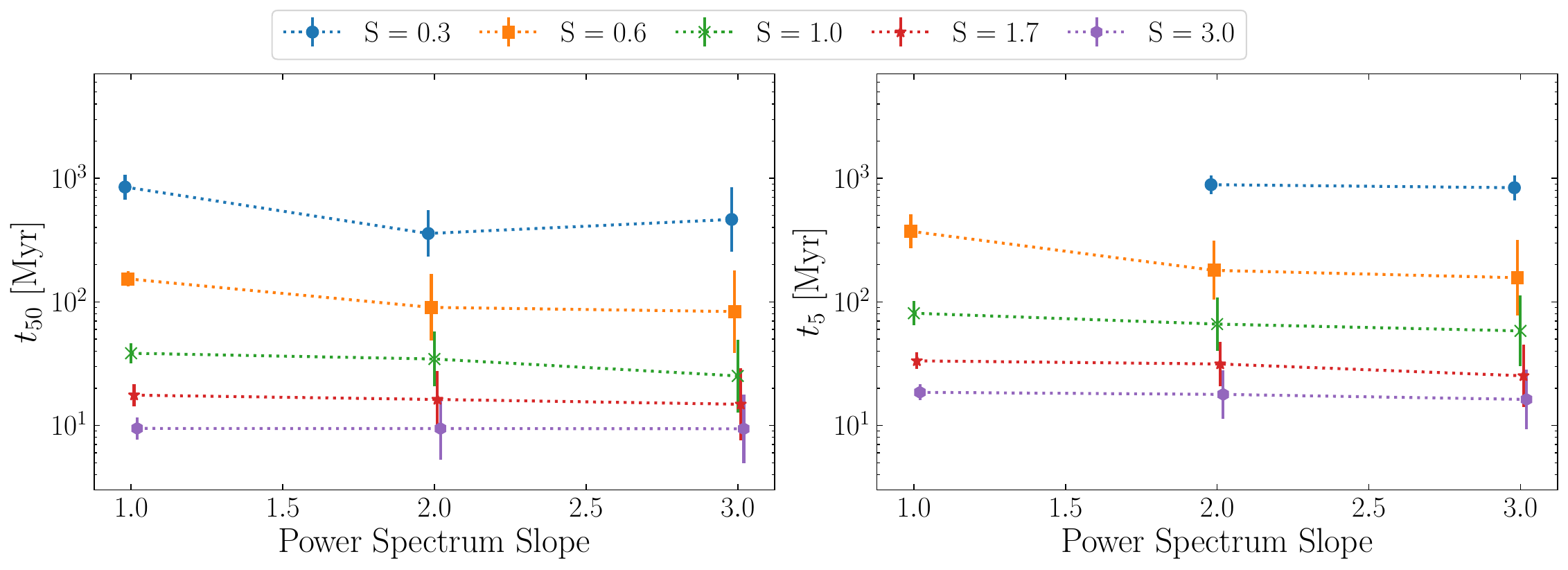}
    \caption{The mean $50~$per cent (\textit{left panel}) and $5~$per cent (\textit{right panel}) dissolution time-scales, i.e.\ the time at which these respective percentages of the initial cluster mass remain, as a function of power spectrum slope for tidal history normalisation factors of $S=\{0.3, 0.6, 1, 1.7, 3\}$. For each data point, the error bar demonstrates the standard deviation in the dissolution time-scale across the 20 corresponding realisations of the tidal history.}
    \label{fig:tdissplot}
\end{figure*}

In order to quantify how the uniqueness of a cluster's tidal history impacts its evolution, we first determine the $50~$per cent and $5~$per cent dissolution time-scales, i.e.\ the time at which these respective percentages of the initial cluster mass remain. We examine the fiducial clusters (clusters above the horizontal line in Table \ref{table:models}) evolved with tidal histories generated for the same power spectrum slope and tidal history normalisation. Figure \ref{fig:tdissplot} illustrates the mean dissolution time-scales of all models with tidal histories generated from power spectra of the same slope and using the same normalisation factor, as a function of power spectrum slope. Note that the N50\_R8\_A1\_S03 models are not featured in the left panel of Figure \ref{fig:tdissplot}, because none of these models reach $5~$per cent of the initial cluster mass within the $5~\gyr$ tidal histories that were generated. 

Figure \ref{fig:tdissplot} demonstrates that the $50~$per cent and $5~$per cent dissolution time-scales of clusters are nearly independent of the power spectrum slope used to generate its tidal history. However, these time-scales depend strongly on the tidal history normalisation. For different normalisations, the dissolution time-scales are generally separated by more than the typical error bar, which reflects the standard deviation due to different realisations of the tidal history from the same power spectrum. This result shows that the impact of the tidal history normalisation on the dissolution time-scale is much stronger than that of the power spectrum slope or of the details of the specific realisation of the tidal history.

The N50\_R8\_A1\_S03 and N50\_R8\_A1\_S06 models are the only cases where the $50~$per cent dissolution time-scales appear to have some minor dependence on power spectrum slope, as their dissolution times are nearly double those of models with slopes of $\alpha=2$ and $\alpha=3$ and with the same normalisations. Even within the standard deviation of the N50\_R8\_A1\_S03 and N50\_R8\_A1\_S06 models, the dissolution time-scales are not comparable. In fact, for these two models, the standard deviations in dissolution time-scales are very small. Such low standard deviations suggests that the effects of high frequency shocks are minimal when the shock strength is weak, such that the clusters go through periods of evolving in a near constant background tidal field and not responding to the weak and short tidal shocks.

\begin{figure*}
    \centering
    \includegraphics[width=\hsize]{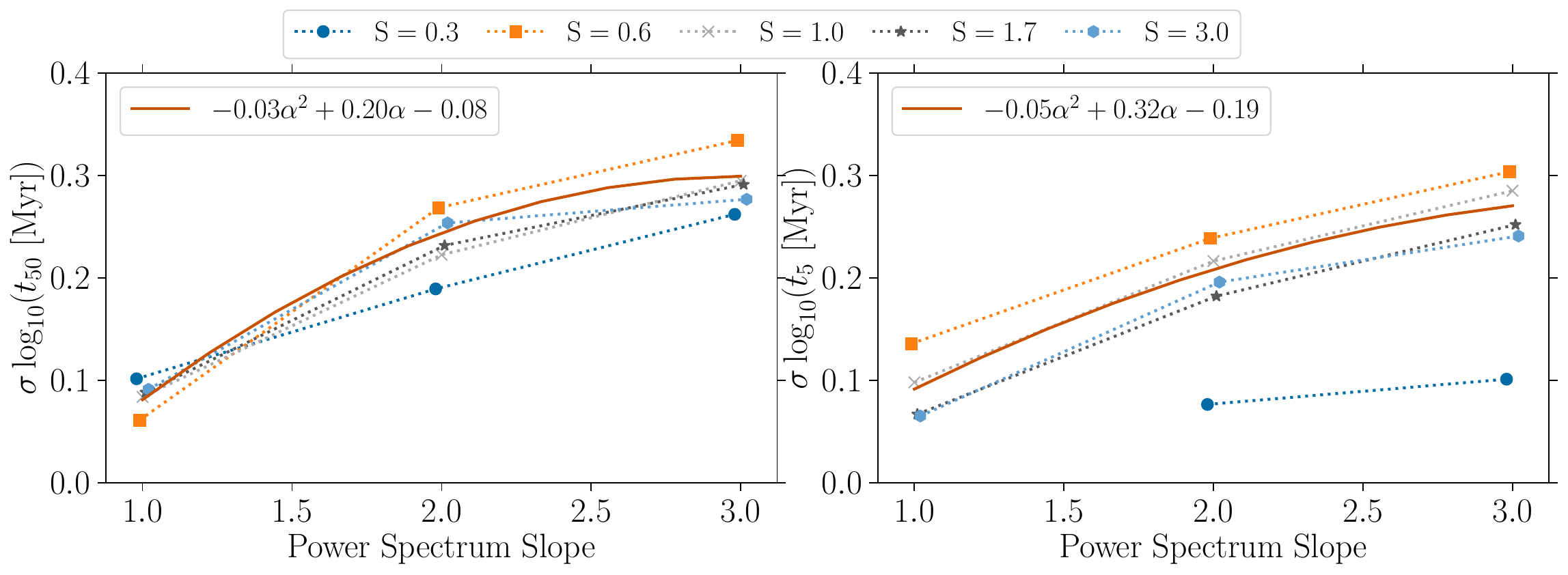}
    \caption{Dispersion in the $\log_{10}$ of the $50~$per cent (\textit{left panel}) and $5~$per cent (\textit{right panel}) dissolution time-scales, i.e.\ the time at which these respective percentages of the initial cluster mass remain, as a function of power spectrum slope for tidal history normalisation factors of $S=\{0.3, 0.6, 1, 1.7, 3\}$. Solid lines correspond to a quadratic fit to models with normalisation factors $S>0.3$.}
    \label{fig:sigtdissplot}
\end{figure*}

Figure \ref{fig:sigtdissplot} takes a closer look at the standard deviation in the logarithm of the dissolution time-scales amongst model clusters with tidal histories generated from power spectra with the same slope and using the same normalisation, plotted as a function of power spectrum slope. Focusing first on the $50~$per cent dissolution time-scales, there is a clear dependence of the standard deviation on power spectrum slope. Tidal histories generated from power spectra with steeper slopes ($\alpha = 3$) yield larger standard deviations in the $50~$per cent dissolution time of clusters than when smaller slopes are used ($\alpha = 1$). Physically, this result means that tidal histories characterised by higher-frequency shocks result in more similar dissolution time-scales, which is consistent with the similarity in mass loss histories identified in Figure~\ref{fig:mplot}. For a given slope, there appears to be no dependence of the standard deviation of the dissolution time-scale on the normalisation of the tidal history.

The relationship between the standard deviation in the logarithm of the $50~$per cent dissolution times, $\sigma[\log_{10}(t_{50})]$, and the slope of the power spectrum $\alpha$ is well-described by a quadratic function of the form:
\begin{equation}
    \label{eq:sigt}
    \sigma[\log_{10}(t_{50})]=a \alpha^2 +b \alpha + c
\end{equation}
with $a=-0.05 \pm 0.01$, $b=0.32 \pm 0.05$, and $c=-0.19 \pm 0.05$. Therefore, we can now approximate the range of dissolution time-scales that any given cluster may have if subjected to different tidal histories generated from the same power spectrum. This relation suggests that the evolution of entire star cluster populations in cosmologically-motivated potentials may potentially be approximated without the need to directly simulate each cluster.

A similar trend is found in the right panel of Figure~\ref{fig:sigtdissplot} for the relationship between $\sigma[\log_{10}(t_{5})]$ and the slope of the power spectrum, with the exception of models with tidal history normalisation factors of $S=0.3$. Models in which the tidal history has been scaled by a factor $S=0.3$ show very little variation in $\sigma[\log_{10}(t_{5})]$, because the normalisation reduces the ability of individual shocks to inject energy into the cluster. The inner regions of the clusters are especially unaffected by these weak shocks. As a result, their evolution is comparable to being subject to a weak, static external tidal field, with shocks no longer being the dominant source of cluster dissolution. Minimising the effectiveness of shocks in turn minimises the uniqueness of individual cluster tidal histories and suppresses the standard deviation of the dissolution time-scale. A slightly higher standard deviation is observed for higher power spectrum slopes ($\alpha = 3$), as low-frequency shocks result in clusters spending some time in a slightly stronger or weaker static tidal field.

Focusing on models with tidal history normalisation factors $S>0.3$, we fit the relationship between $\sigma[\log_{10}(t_{5})]$ and slope $\alpha$ with the quadratic function from equation~(\ref{eq:sigt}). We find that the best-fitting parameters are $a= -0.03 \pm 0.02$, $b=0.20 \pm 0.07$, and $c= -0.08 \pm 0.06$. The uncertainties in the quadratic fits to $\sigma[\log_{10}(t_{5})] (\alpha)$  are slightly larger than those in the fit to $\sigma[\log_{10}(t_{50})] (\alpha)$. 
This contrast is driven by the fact that differences in tidal history normalisation (which are not included in the fit) more strongly affect $\sigma[\log_{10}(t_{5})]$ than $\sigma[\log_{10}(t_{50})]$. Despite this slightly different sensitivity of the standard deviation of the dissolution time-scale, the mean ratio of ${t_5}/{t_{50}} = 2.0 \pm 0.6$ for all models, independently of the power spectrum slope and tidal history normalisation. Finally, Figure~\ref{fig:sigtdissplot} also shows that $\sigma[\log_{10}(t_{50})]$ is generally larger than $\sigma[\log_{10}(t_{5})]$. This is the direct result of the fact that we consider the standard deviation in the logarithm of the dissolution time-scale (which represents a relative standard deviation). Because $t_5$ is obtained by integrating over a larger part of the tidal history than $t_{50}$, we find smaller relative standard deviations.

\subsection{Evolution of cluster mass and density} \label{sec:density}

\begin{figure*}
    \includegraphics[width=0.98\textwidth]{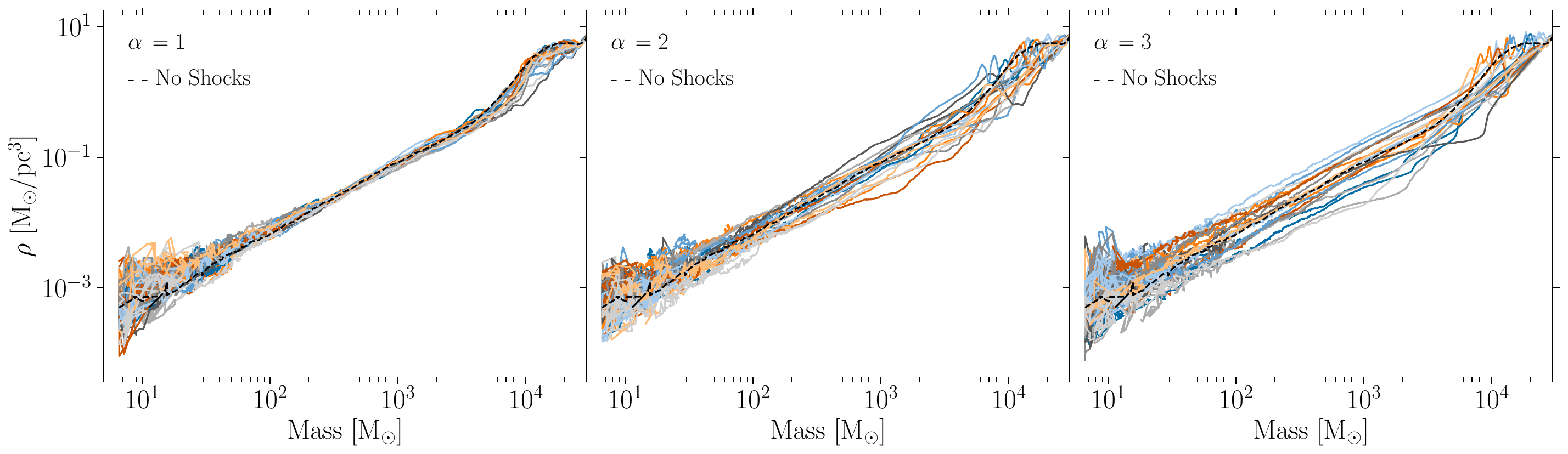}
    \caption{Cluster density within the half-mass radius as a function of its remaining mass for power spectrum slopes of $\alpha=1$ (\textit{left panel}), $\alpha=2$ (\textit{middle panel}), and $\alpha=3$ (\textit{right panel}). All tidal histories have a normalisation of $S=1$, and each coloured line corresponds to the evolution of an individual cluster until its dissolution. For comparison, we include in each panel the evolution of a model cluster subject to a static tidal field of strength equal to the median tidal field strength of the 20 realisations (dashed black line).}
    \label{fig:pplot}
\end{figure*}

Next, we consider how the evolution of a cluster's structure, namely its mass and density, differ when subjected to tidal histories generated from the same power spectrum. For the mass and density calculation, we only consider stars that are gravitationally bound to the cluster. As an example, Figure \ref{fig:pplot} demonstrates how the density within the half-mass radius evolves as a function of cluster mass for 20 star clusters with tidal histories generated from power spectra with different slopes and using the same tidal history normalisation of $S=1$. For a given power spectrum, the evolution of cluster density and mass is extremely similar between individual models, down to remaining mass fractions of $1~$per cent, despite the fact that each cluster is subjected to a unique tidal history. There is also very little difference between the evolution of models that experience non-static and static tidal histories, the latter of which is shown for comparison purposes in Figure \ref{fig:pplot}.

\begin{figure}
    \centering
    \includegraphics[width=0.48\textwidth]{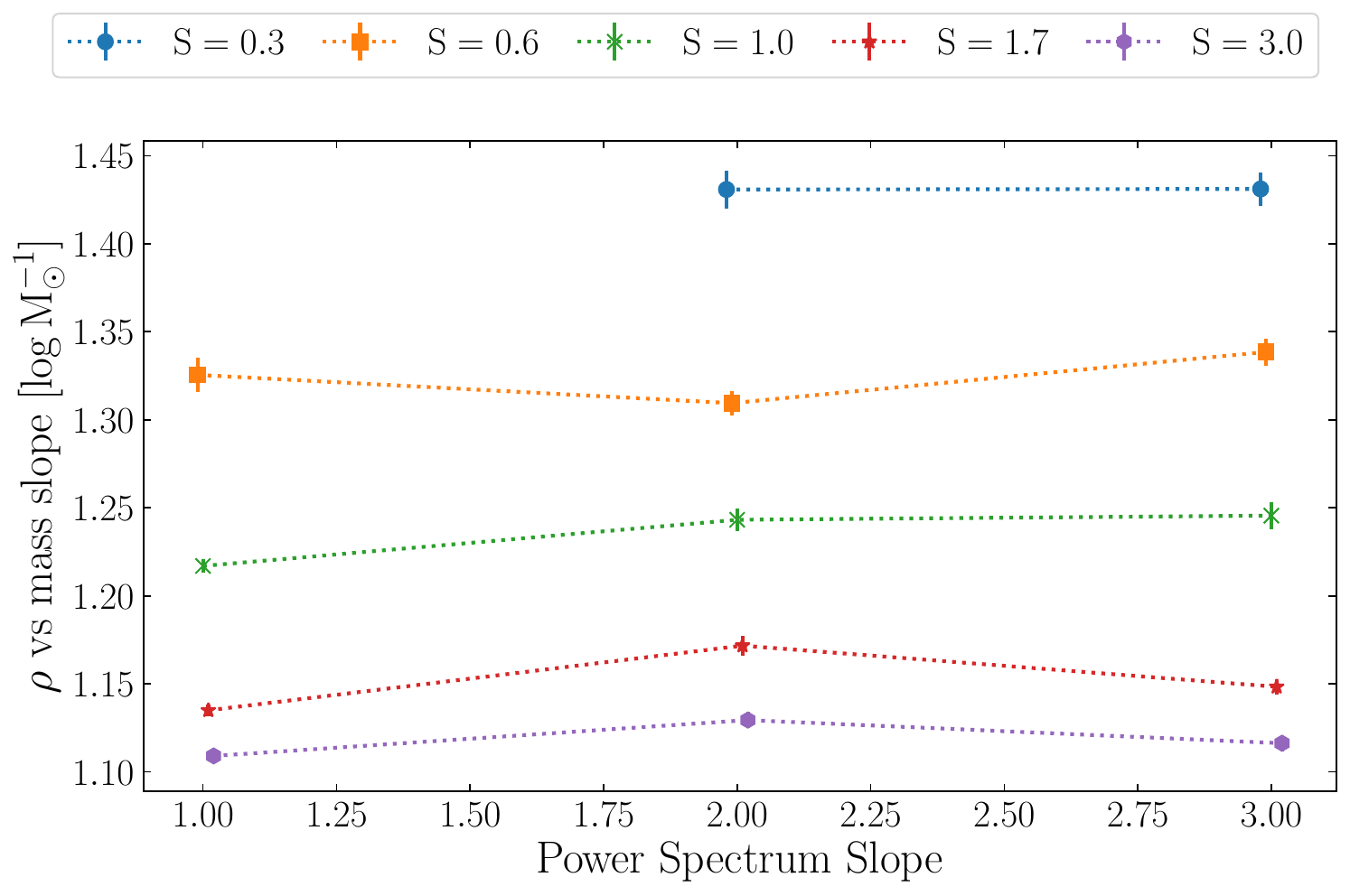}
    \caption{Best-fitting power-law slope to the relationship between cluster density and mass, shown as a function of power spectrum slope for tidal history normalisation factors of $S=\{0.3, 0.6, 1, 1.7, 3\}$.}
    \label{fig:pslopeplot}
\end{figure}

To investigate and compare the co-evolution of cluster density and mass for our entire suite of simulations, we first determine the best-fitting power-law slope of the $\log \rho$--$\log M$ relation for each power spectrum slope. In the fit, we use all 20 models evolved under tidal histories generated from a power spectrum with a given slope and using a given tidal history normalisation factor. The models are first binned by $\log M$, and the mean value of $\log \rho$ in a given $\log M$ bin is then used to establish the mean relationship. A linear fit is then made to the binned data, with the slope of the fit shown in Figure \ref{fig:pslopeplot}. Across all of the models, the density-mass relationship is consistent with $\rho \propto M^{1.10{-}1.45}$. This relationship implies a mass-radius relationship $r_{\rm h}\propto M^\beta$ with the slope $\beta$ ranging from $\beta=-0.15$ to $\beta=-0.03$. In other words, the radius is nearly constant, with a slight expansion as the mass decreases. This expansion is likely driven by the energy injection from the tidal shocks, but weakens for stronger shocks, because a larger fraction of the accelerated stars are gravitationally unbound immediately. If our simulations had also included a static background tidal field, the shock-driven expansion may not have been noticeable, as accelerated stars would have been stripped off by the tidal field.

It is important to note that the N50\_R8\_A1\_S03 models are not featured in Figure \ref{fig:pslopeplot}, because the tidal field is so weak that clusters do not decrease in density as they lose mass. Instead, clusters first lose mass at a constant density, which is consistent with mass loss in a static tidal field. As cluster approach dissolution, they increase in density as the outer layers are stripped and the half-mass radius moves inwards. Furthermore, given that clusters evolve for several initial relaxation times, the core is able to become denser over time since the tidal field is too weak to affect the inner regions of the clusters.

Figure \ref{fig:pslopeplot} demonstrates that the mass-density relationship at a given tidal history normalisation is independent of the power spectrum slope used to generate the cluster's tidal history. Hence, the evolution of cluster density is independent of the tidal shock frequency that the clusters experience. The evolution primarily depends on the total energy injected into the cluster. We find that the power-law relationship between density and mass is steeper for lower normalisations, independently of the power spectrum slope. This result is expected, given that stronger tidal fields (i.e.\ higher normalisation) are more capable of energizing stars in the inner regions of the cluster, which in turn can cause the cluster's core to expand and thus its density to decrease further.

\subsection{Dependence on the cluster properties} \label{sec:mass}

\begin{figure*}
    \centering
    \includegraphics[width=\hsize]{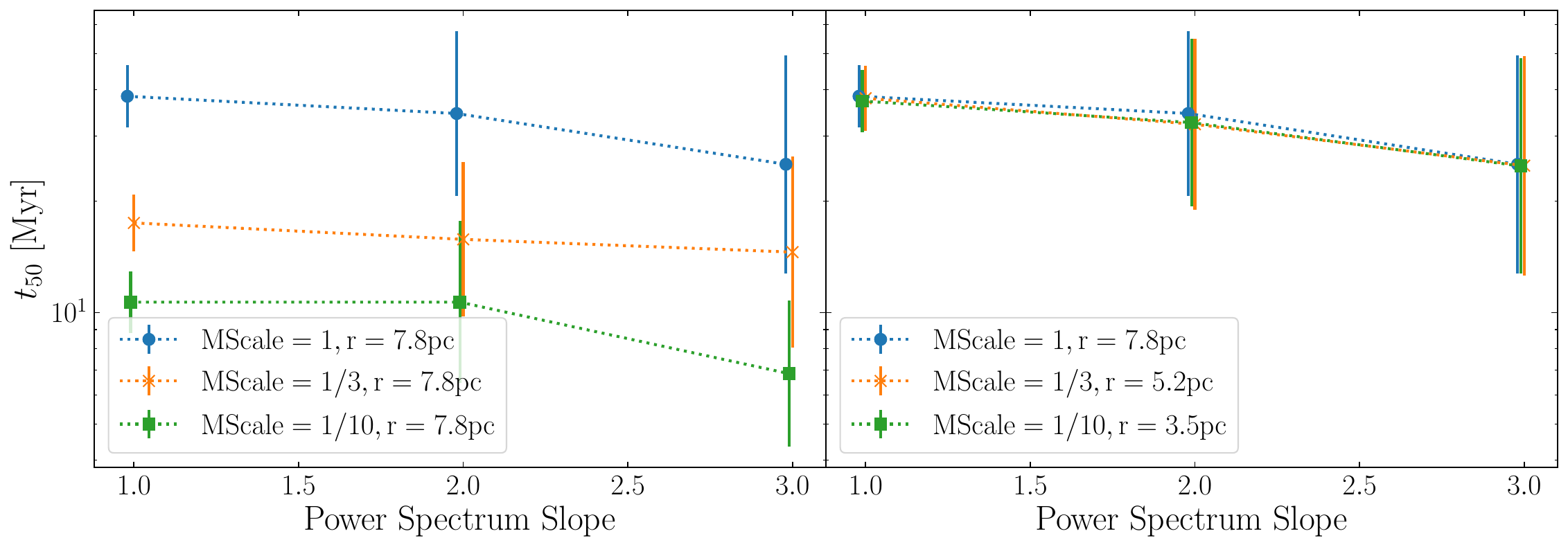}
    \caption{Relation between the $50~$per cent dissolution time and the power spectrum slope used to generate the tidal histories, for clusters with different initial masses but similar radii (\textit{left panel}) and different initial masses and radii but similar densities (\textit{right panel}). All of these clusters experience tidal histories with a normalisation factor of $S=1$. For each data point, the error bar demonstrates the standard deviation in the dissolution time-scale across the 20 corresponding realisations of the tidal history.}
    \label{fig:tdissmplot}
\end{figure*}

In the previous sections, all of our clusters start with identical initial half-mass radii and masses. These simulations suggest that a cluster's dissolution time and its mass-density evolution are independent of the power spectrum slope used to generate its tidal history. Here we explore how these conclusions may depend on the initial density of the cluster. In Figure \ref{fig:tdissmplot}, we show the $50~$per cent dissolution time as a function of the slope of the power spectrum for model clusters with different masses, sizes, and densities than our fiducial models (see the second half of Table~\ref{table:models}). All of these clusters experience tidal histories with a  normalisation of $S=1$.

\begin{figure*}
    \centering
    \includegraphics[width=\hsize]{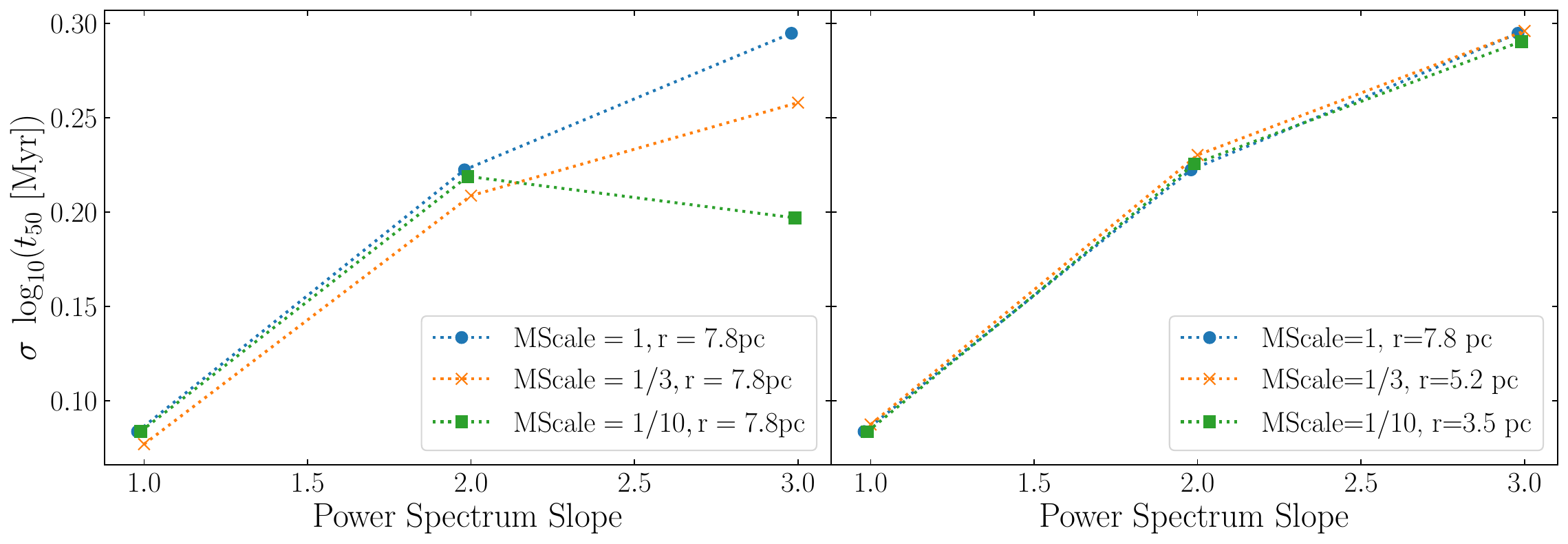}
    \caption{Dispersion in the $\log_{10}$ of the $50~$per cent dissolution time-scales, i.e.\ the time at which these respective percentages of the initial cluster mass remain, as a function of power spectrum slope for clusters with different initial masses but similar radii (\textit{left panel}) and different initial masses and radii but similar densities (\textit{right panel}). All of these clusters experience tidal histories with a normalisation factor of $S=1$. For each data point, the error bar demonstrates the standard deviation in the dissolution time-scale across the 20 corresponding realisations of the tidal history.}
    \label{fig:tdissmplot_sig}
\end{figure*}

In the left panel of Figure \ref{fig:tdissmplot}, we add simulations with initial cluster masses that are $1/3$ and $1/10$ of our fiducial model N50\_R8\_A1\_S10. Their initial radii are kept the same as for model N50\_R8\_A1\_S10, such that their densities are lower. As expected, the dissolution times are shorter for the lower density models, and we find that the $50~$per cent dissolution time is still largely independent of the power spectrum slope used to generate the tidal history. Figure \ref{fig:tdissmplot_sig} further shows that the dispersion in $\log_{10} (t_{50})$ is also very similar for models that have tidal histories generated with the same power spectrum slope. The one exception to the above statements is the models with 1/10 mass scaling, where we see that the mean $50~$per cent dissolution time and $\sigma[\log_{10} (t_{50})]$ are smaller for $\alpha=3$ than for $\alpha=1$ or $\alpha=2$. While the $50~$per cent dissolution times still agree within $1\sigma$, we find that low frequency shocks are capable of accelerating the dissolution of very low density clusters. These clusters are the most tidally filling in our dataset, so their dissolution carries a stronger contribution from the mean background tidal field that is generated by the slowest shocks.

In the right panel of Figure \ref{fig:tdissmplot}, we add simulations with initial masses that are $1/3$ and $1/10$ of our fiducial model N50\_R8\_A1\_S10, as before. This time, the initial radii are changed accordingly, to ensure that the cluster's initial density is the same as N50\_R8\_A1\_S10. The dissolution times of all models are nearly identical, demonstrating that a cluster's dissolution time-scale is set by its initial density rather than its mass or radius alone. Similarly in the right panel of Figure \ref{fig:tdissmplot_sig}, the standard deviation of $\log_{10} (t_{50})$ is unchanged for models with identical initial densities and tidal histories generated with the same power spectrum slope.

\begin{figure*}
    \centering
    \includegraphics[width=\hsize]{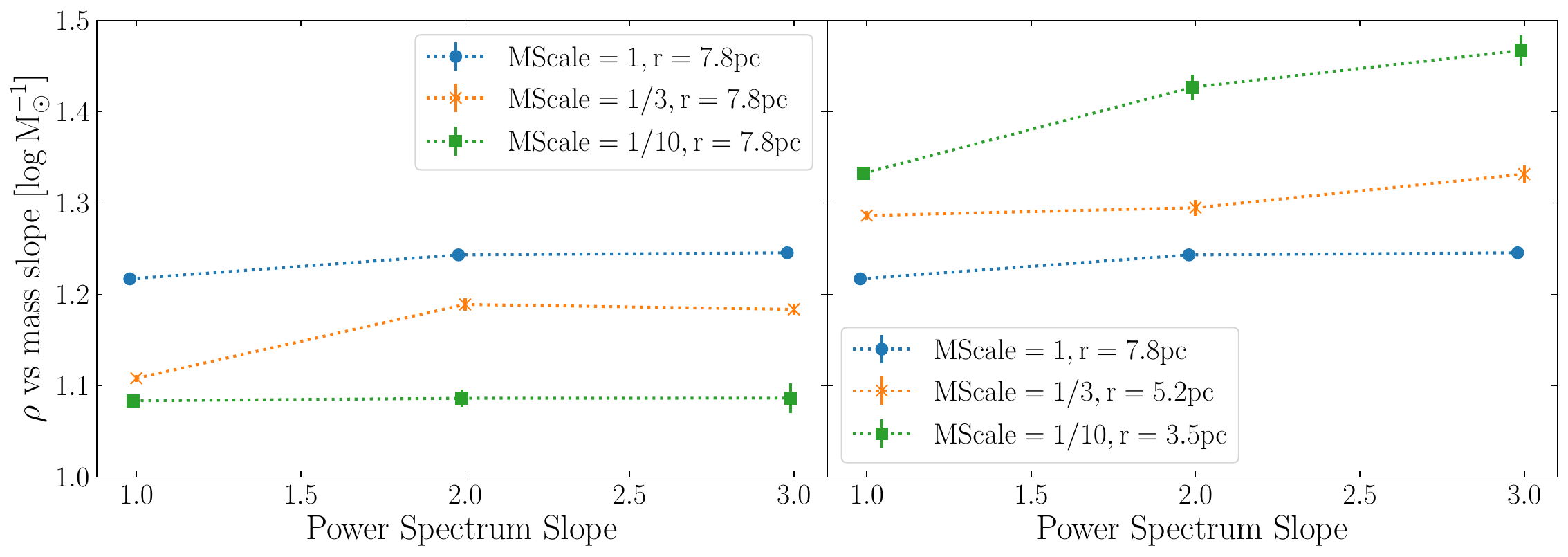}
    \caption{Best-fitting power law slope to the relationship between cluster density and mass, shown as a function of power spectrum slope for clusters with different initial masses but similar radii (different densities, \textit{left panel}) and different initial masses but similar densities (fixed density, \textit{right panel}) and a tidal history normalisation factor of $S=1$.}
    \label{fig:pslopemplot}
\end{figure*}

Similar to Section \ref{sec:density}, we now explore the relationship between cluster density and mass for clusters with different initial masses, sizes, and densities than our fiducial models. We determine the best-fitting power-law slope of the density-mass relations for all the models for which we modify either their masses or their sizes (second half of Table~\ref{table:models}), and we show them in Figure~\ref{fig:pslopemplot}. As in Figure \ref{fig:tdissmplot}, the left panel of Figure \ref{fig:pslopemplot} compares clusters of different masses and fixed radii (i.e.\ different densities), while the right panel compares clusters of different masses and radii (i.e.\ fixed densities). In all cases, the best-fitting slope of the power-law mass and density relationship decreases as initial density decreases. For fixed initial densities, the best-fitting slope of the power-law mass and density relationship increases as cluster mass and size decrease. We further find that the best-fitting slope of the power-law mass and density relationship depends is not always constant with the power spectrum slope, in contrast with our findings in Figure~\ref{fig:pslopeplot} for fixed cluster properties. More specifically, the intermediate density clusters (yellow crosses in the left panel of Fig.~\ref{fig:pslopemplot}) have a shallower density-mass relationship for shallow power spectra ($\alpha=1$) than for steep power spectra ($\alpha\geq2$). Similarly, high-mass clusters at the fiducial density (green squares in the right panel of Fig.~\ref{fig:pslopemplot}) have a density-mass slope that increases with power spectrum slope, despite having the same density as the N50\_R8\_A1\_S10 fiducial model. Nonetheless, these differences in the slope of the density-mass relationship as a function of initial cluster properties are small; the slopes fall within the same range as that spanned when changing the tidal history normalisation factor (Figure~\ref{fig:pslopeplot}). As a result, across all models we conclude that the density-mass relationship is consistent with $\rho \propto M^{1.08-1.45}$, independently of the slope of the power spectrum, the normalisation of the tidal history, cluster mass, cluster size, or cluster density.

\section{Discussion and Conclusions} \label{sec:conclusion}

We have explored how star cluster evolution depends on the properties of its time-variable tidal history by running a large suite of 540 direct $N$-body simulations of star cluster disruption. Specifically, we have investigated whether the statistical properties of a cluster's tidal history determine its evolution or whether individual tidal perturbations must be described independently. To systematically study the dependence of cluster evolution on the tidal history, we generate tidal tensors from power-law power spectra in Fourier space. Each tidal history depends solely on two parameters: the slope of the power spectrum and the normalisation factor by which the tidal history is multiplied to control the rate of star cluster disruption. For each tidal history, we only generate the $T_{xx}$ component of the tensor, with all other components set to zero. We also only consider extensive shocks, such that $T_{xx}\geq0$ at all times.

For a given combination of power spectrum slope $\alpha$ and tidal history normalisation factor $S$, 20 star cluster simulations are performed with tidal histories generated as different random realisations from the same power spectrum. This determines the range of evolutionary paths a cluster can take for statistically indistinguishable tidal histories. We find that the mean $50~$per cent and $5~$per cent dissolution time-scales, i.e.\ the times at which these respective percentages of the initial cluster mass remain, are independent of the power spectrum slope $\alpha$. This shows that a cluster's dissolution time-scale primarily depends on the strength of the shocks it experiences and thus the total amount of energy injected, rather than the frequency of the tidal shocks. The mean dissolution times remain unchanged for clusters with similar densities, albeit with different masses and sizes. However, consistent with a wide body of literature on tidal shock-driven cluster disruption, we find that the dissolution time-scale decreases with increasing initial cluster density for a given tidal history.

By contrast to the absent correlation between power spectrum slope and the absolute dissolution time-scale, we find that the standard deviation of the $50~$per cent and $5~$per cent dissolution time-scales exhibits a strong dependence on $\alpha$. For low values of $\alpha$, which corresponds to shallow power spectra and high-frequency shocks, the standard deviation in cluster dissolution time-scales is low ($\sim0.1$~dex). Conversely, for higher values of $\alpha$, which corresponds to steep power spectra and low-frequency shocks, cluster dissolution times have standard deviations up to $\sim0.3$~dex. This finding suggests that the timing of strong, long-period shocks can cause major differences in a cluster's evolution. However, when clusters are subjected to a large number of high-frequency shocks, their evolution is generally quite similar.

The standard deviation in $\log_{10}(t_5)$ and $\log_{10}(t_{50})$ increases with the power spectrum slope $\alpha$ in a way that is well-fitted by a quadratic function. The exact relationships are (see Section \ref{sec:results} for details):
\begin{equation}
   \sigma [\log_{10}(t_{50})] = (-0.05 \pm 0.01) \alpha^2 + (0.32 \pm 0.05) \alpha - (0.19 \pm 0.05) ,
\end{equation}
and
\begin{equation}
   \sigma [\log_{10}(t_{5})] = (-0.03 \pm 0.02) \alpha^2 + (0.20 \pm 0.07) \alpha - (0.08 \pm 0.06) .
\end{equation}
Note that models with normalisation parameters of $S=1/3$ do not follow these relations, as the low normalisation results in the strength of individual shocks being too low to significantly alter the cluster's density profile. In physical units, this normalisation factor corresponds to mean tidal field strengths of $\overline{T} \sim 0.007 Myr^{-2}$. Our approximations for the standard deviation of the dissolution time-scales should therefore only be used to describe tidal histories with mean tidal field strengths $\overline{T} > 0.007 Myr^{-2}$.

The fact that the above relations exist opens up the possibility of being able to sample cluster evolutionary tracks for a given statistical representation of the tidal history. Given the high computational cost of direct $N$-body star cluster simulations, either performed within a cosmological simulation or separately given a tidal history, it will be extremely advantageous to instead be able to sample a cluster's evolution from a well constrained distribution. 

In addition to exploring each cluster's mass loss history we have also considered the evolution of cluster half-mass density $\rho$. While clusters initially lose mass while keeping their density fixed, mass loss due to shocks results in a mass-size evolution that can be characterized by $\rho \propto M^{1.08-1.45}$. This range of slopes reflects a weak dependence on the tidal history normalisation factor $S$ (which we vary by an order of magnitude), and is independent of the power spectrum slope $\alpha$. For a fixed initial cluster density, tidal histories with larger normalisation factors (stronger tidal shocks) have shallower mass-density relationships than histories with lower normalisation factors (weaker tidal shocks). However, changing the cluster's initial mass, size, or density leads to mass-density relationships in our models that can either be steeper or shallower. Unfortunately, this behaviour seems to be stochastic. The lack of a predictable relationship between a cluster's mass-size evolution and the initial density for the models considered here negates the possibility of sampling from a well-constrained distribution function for a given tidal history, which can be done for a cluster's mass loss history. However, knowing that the power-law relationship has a slope between $1.08$ and $1.45$ still allows for constraints to be placed on the cluster's evolutionary pathway. 

The simulations presented in this work mark a first step towards being able to approximate the evolution of an entire population of star clusters using the statistical properties of their tidal histories. Our results indicate that a Fourier analysis of a given tidal history may be all that is required to establish the range of possible mass loss histories and density evolutionary tracks of a cluster that experiences such a tidal history. Due to the simple, single-component nature of the adopted tidal tensors, we must acknowledge the possibility that a statistical description of a star cluster's evolution may only be possible with such simplified tidal histories. If the effects of tidal heating on cluster evolution are primarily driven by the time evolution of a single global parameter of the tensor (e.g.\ the maximum eigenvalue as a function of time), then the conclusions drawn based on the simple tidal histories used here might still apply more generally. For future applications of our new approach, it remains desirable to extend our analysis to tidal histories defined in three dimensions, rather than only for the first component of the tidal tensor. Furthermore, our exploration of the dependence on initial cluster masses and densities would benefit from a finer sampling of parameter space, beyond the 540 simulations that were used here. Finally, it will be necessary to connect the power spectrum slope to the physical properties of the perturber, which most commonly will correspond to the cold ISM (see the discussion in Section~\ref{sec:intro}).

While the results of our study imply the effects of tidal heating on cluster evolution could be approximated based on a Fourier analysis of its tidal history, it does not take into account other factors known to govern cluster evolution. Stellar evolution is known to affect early cluster evolution and relaxation is capable of affecting the evolution of clusters with longer dissolution times \citep{heggie03}. While we expect tidal shocks to be the dominant external mass loss mechanism experienced by clusters in their natal galactic disk, the shock-driven mass loss rate could be influenced by these other mass loss mechanisms. Any possible forms of covariance between these mechanisms should be explored too.

The above future steps will allow sampling the evolution of entire star cluster populations evolving in unique galactic environments from statistically representative evolutionary tracks. Minimizing the need to model the evolution of each individual cluster in detailed galaxy models will improve our ability to make use of clusters as tracers of galaxy formation, evolution, and structure.

\section*{Acknowledgements}
JW acknowledges computing support and funding provided from Jo Bovy through the  Natural Sciences and Engineering Research Council of Canada (funding reference number RGPIN-2020-04712). MRC gratefully acknowledges the Canadian Institute for Theoretical Astrophysics (CITA) National Fellowship for partial support; this work was supported by the Natural Sciences and Engineering Research Council of Canada (NSERC). JMDK gratefully acknowledges funding from the European Research Council (ERC) under the European Union's Horizon 2020 research and innovation programme via the ERC Starting Grant MUSTANG (grant agreement number 714907). COOL Research DAO is a Decentralised Autonomous Organisation supporting research in astrophysics aimed at uncovering our cosmic origins.  This work was made possible in part by the facilities of the Shared Hierarchical Academic Research Computing Network (SHARCNET: \url{www.sharcnet.ca}) and Compute/Calcul Canada.


\bibliography{bibfile}{}
\bibliographystyle{aasjournal}



\end{document}